\documentclass{moriond}

\def\Journal#1#2#3#4{{#1} {\bf #2}, #3 (#4)}

\def\be{\begin{equation}}
\def\ee{\end{equation}}
\def\bea{\begin{eqnarray}}
\def\eea{\end{eqnarray}}

\newcommand{\Photo}{}

\begin{document}
\vspace*{4cm}
\title{DENSE QCD(-LIKE) MATTER IN STRONG MAGNETIC FIELDS}

\author{HELENA KOLE\v{S}OV\'{A}}

\address{Department of Mathematics and Physics, University of Stavanger\\
Kj{\o}lv Egeland building, 4021 Stavanger, Norway}

\maketitle\abstracts{
As recently shown, the ground state of dense QCD matter in a strong magnetic field is an inhomogeneous condensate of neutral pions which was named Chiral Soliton Lattice (CSL) phase in analogy with a state present for chiral magnets. We continue this work by showing that similar phase is present for a class of QCD-like theories which are accessible to standard lattice Monte-Carlo simulations unlike the QCD itself which suffers from the notorious sign problem if non-zero baryon chemical potential is considered. On the occasion of the Gravitation session of the Moriond conference we also comment on the possible relevance of the CSL phase for neutron stars and gravitational wave observations.}

%\section{Introduction}
%QCD at low energies as a strongly interacting theory can not be approached by standard perturbation theory, however, in many cases the first principle results can be provided by lattice simulations. This is, unfortunately, not the case if non-zero baryon chemical potential is considered since the so-called sign problem is present, i.e., the non-positivity of the Dirac determinant does not allow for using standard Monte-Carlo methods. Theoretical studies of the QCD phase diagram are then limited to model calculations unless the energies are so low or so high that the chiral perturbation theory or perturbative QCD applies, respectively.

%On the experiment side, QCD phase diagram is explored in heavy ion collisions and also neutron stars give us an valueable information about dense QCD matter. In both these cases considering the presence of the magnetic field is relevant. Dense QCD matter in strong magnetic fields was studied by means of the chiral perturbation theory~\cite{Son:2007ny,Brauner:2016pko} and for strong enough magnetic fields a new phase of matter was found as described in section~\ref{secCSL}. Interestingly, the baryon densities reached in neutron stars are accessible by the chiral perturbation theory predictions although the magnetic field necessary for the appearance of this phase are too strong to be present even in the highly magnetized neutron stars, the so-called magnetars. We discuss this in section~\ref{secNeutronStars}.

\section{Chiral Soliton Lattice Phase}
\subsection{QCD}\label{secCSL}
Dense QCD matter was recently studied~\cite{Son:2007ny,Brauner:2016pko} by means of the chiral perturbation theory with two light flavours when % are assumed, the degrees of freedom in the chiral perturbation theory are the three pions with electric charges $\pm 1$ and $0$. 
large constant background magnetic field is present. In such a setting the charged pions attain large effective masses and only the neutral pion remains relevant as a low-energy degree of freedom. Due to the chiral anomaly, the neutral pion field couples to the magnetic field and baryon chemical potential. The corresponding negative contribution to the Hamiltonian density makes the configurations with non-zero gradient of the neutral pion field energetically preferable if the magnetic field $H$ and baryon chemical potential $\mu$ reach the values as large as~\cite{Brauner:2016pko}
\begin{equation}\label{boundCSL}
\mu H \geq 16 \pi m_\pi f_\pi^2
\end{equation}
(here $m_\pi$ and $f_\pi$ are pion mass and decay constant, respectively). Interestingly, the ground state configuration then carries non-zero baryon number density $n_B$ proportional to the gradient of the pion field and it can be shown~\cite{Brauner:2016pko} that for baryon chemical potential within the reach of chiral perturbation theory, $n_B$ reaches few-times nuclear saturation density, i.e., the values relevant for the cores of neutron stars.

\subsection{QCD-like theories}
Unfortunately, QCD with non-zero baryon chemical potential can not be approached by standard lattice Monte-Carlo methods due to the presence of the so-called sign problem, i.e., due to the non-positivity of the Dirac determinant. On the other hand, it is known that for theories with quarks in real or pseudo-real representations of the gauge group (an example of such a theory is the so-called two-color QCD based on $SU(2)$ gauge group with quarks in fundamental representation) the Dirac determinant is positive even for non-zero baryon chemical potential. This remains to be true also if strong magnetic fields are considered provided the charges of the $u$- and $d$-quarks satisfy~\cite{Brauner:2019rjg} $q_u = -q_d$.

This class of QCD-like theories was studied~\cite{Brauner:2019rjg} using effective field theory methods. In general, the coset spaces $SU(4)/SO(4)$ and $SU(4)/Sp(4)$ corresponding to 9 and 5 pseudo-Goldstone bosons have to be considered in the real and pseudo-real case, respectively; however, in strong magnetic fields, only three neutral pseudo-Goldstone bosons remain light in both cases, namely, the neutral pion and a neutral diquark-antidiquark pair. The real and pseudo-real theories can be, hence, studied at once.

%This can be seen also from the fact that if only the subgroups commuting with the electric charge generator $Q$ are considered, then the symmetry breaking pattern is $SU(2)\times SU(2)\times U(1)_Q/SU(2)_{\mathrm{diag}}\times U(1)_Q$ in both real and pseudo-real case. Although this coset space is different from the coset space relevant for QCD (the baryon number generator is part of $SU(2)_{\mathrm{diag}}$ here), the isomorphism with this coset space allows for simple derivation of the Wess-Zumino-Witten term capturing the effect of the chiral anomaly which is in turn crucial for appearance of a CSL-like phase. 

The Lagrangian including the Wess-Zumino-Witten term capturing the effect of the chiral anomaly was derived~\cite{Brauner:2018zwr,Brauner:2019rjg}, and the analysis of the corresponding Hamiltonian density suggested that two non-trivial ground states appear. First, for $\mu>m_\pi$ and low magnetic fields a Bose-Einstein condensate of diquarks appears. Second, above certain critical magnetic field satisfying
\begin{equation}\label{HcritQCDlike}
H_{\mathrm{cr}}\geq \frac{8\pi^2 f_\pi^2}{d}
\end{equation}
(for quarks transforming in a $d$-dimensional representation of the color gauge group) the ground state is a CSL-like condensate of neutral pions. Let us note that for small $d$ (e.g., as in case of two-color QCD), this magnetic field may be rather large and corrections to $f_\pi$ due to the presence of magnetic field have to be considered.
In this case, our study would have to be complemented by an input (possibly by lattice simulations) on the dependence of $f_\pi$ on the magnetic field in order to see if the magnetic fields satisfying Eq.~\ref{HcritQCDlike} can be reached.

\section{Conclusions and outlook: Relevance for the neutron stars and GW observations}\label{secNeutronStars}
A new phase of dense QCD matter present in strong magnetic fields was described and it was argued that similar phase can be seen also in a class of QCD-like theories which are accessible to lattice simulations. Could, however, the CSL phase be observed in nature? As mentioned above, the baryon densities relevant for neutron stars can be reached with moderate chemical potentials, however, Eq.~\ref{boundCSL} implies that the magnetic field $\sim 10^{19}$ G (for $\mu \sim 900$ MeV) is necessary. It was suggested~\cite{Ferrer:2010wz}
that such magnetic fields could be reached in the cores of magnetars; on the other hand, numerical studies solving the full Einstein-Maxwell equations show that for magnetic fields $\sim 10^{18}$ G the neutron star is getting a toroidal shape and static configurations do not exist any more~\cite{%Bocquet:1995je,
Cardall:2000bs}. On the other hand, e.g., the toroidal component of the magnetic field is not taken into account in this kind of works or perhaps considering different equations of state (EoS) might change the results. 

The EoS for the CSL phase can be easily derived using existing results~\cite{Brauner:2016pko} and our preliminary investigations show that this EoS is rather stiff. However, a detailed analysis of a realistic setting (and the question if such a setting exists at all) will be a subject of further work.

%Imagining for a while that high enougn magnetic fields can be reached, it is easy to derive the EoS for the CSL phase using the results of \cite{Brauner:2016pko} and the TOV equations can be solved. This simple exercise shows that the CSL EoS is rather stiff. However, in realistic studies several additional aspects would have to be taken into account. First, in strong magnetic fields the TOV equations do not apply any more since the system can not be considered as spherically symmetric. Second, additional particles would have to be added in order to satisfy the constraints of charge and beta equilibrium. Third, the profile of the magnetic throughout the star would be a necessary input.

As for the possible GW observations, we do not expect to find the signatures of the CSL phase in mergers of neutron stars since magnetars are found to be isolated starts. On the other hand, also the neutron stars which do not have a binary companion could be studied by GW experiments, in particular, due to the possible continuous GW signals~\cite{Glampedakis:2017nqy}.

%\bibliography{references}
%\bibliographystyle{plain}

%\begin{thebibliography}{99}
%\bibitem{ja}C Jarlskog in {\em CP Violation}, ed. C Jarlskog
%(World Scientific, Singapore, 1988).

%\bibitem{ma}L. Maiani, \Journal{\PLB}{62}{183}{1976}.

%\bibitem{bu}J.D. Bjorken and I. Dunietz, \Journal{\PRD}{36}{2109}{1987}.

%\bibitem{bd}C.D. Buchanan {\it et al}, \Journal{\PRD}{45}{4088}{1992}.

%\end{thebibliography}

\end{document}